\documentclass[runningheads]{llncs}

\usepackage{eccv}

\usepackage{eccvabbrv}

\usepackage{graphicx}
\usepackage{booktabs}

\usepackage{xcolor}
\usepackage{booktabs}
\usepackage{multirow}
\usepackage{threeparttable}
\usepackage{graphicx}
\usepackage{makecell}
\usepackage{pifont}
\usepackage{lipsum}
\usepackage{mathrsfs}
\usepackage{bm}
\usepackage{textcomp}
\usepackage{makecell}
\usepackage[linesnumbered,titlenumbered,ruled,vlined,commentsnumbered,noend]{algorithm2e}

\usepackage[accsupp]{axessibility}  

\usepackage[pagebackref,breaklinks,colorlinks,citecolor=eccvblue]{hyperref}

\usepackage{orcidlink}

\newcommand{\figref}[1]{Fig.~\ref{#1}}
\newcommand{\reqref}[1]{Eq.~\eqref{#1}}
\newcommand{\secref}[1]{Sec. ~\ref{#1}}
\newcommand{\tabref}[1]{Table~\ref{#1}}
\usepackage{xspace}
\makeatletter
\DeclareRobustCommand\onedot{\futurelet\@let@token\@onedot}
\def\@onedot{\ifx\@let@token.\else.\null\fi\xspace}
\def\eg{\emph{e.g}\onedot} 
\def\ie{\emph{i.e}\onedot} 
 
\def\etc{\emph{etc}\onedot}

\def\etal{\emph{et al}\onedot}
\makeatother

\begin{document}

\title{Event Trojan: Asynchronous Event-based Backdoor Attacks}

\titlerunning{Event Trojan: Asynchronous Event-based Backdoor Attacks}

\author{Ruofei Wang\inst{1}\orcidlink{0009-0004-3237-9665} \and
Qing Guo\inst{2}\orcidlink{0000-0003-0974-9299} \and
Haoliang Li\inst{3}\orcidlink{0000-0002-8723-8112} \and
Renjie Wan\inst{1}\thanks{Corresponding author.}\orcidlink{0000-0002-0161-0367}}

\authorrunning{Ruofei Wang et al.}

\institute{Department of Computer Science, Hong Kong Baptist University \and
IHPC and CFAR, Agency for Science, Technology and Research (A\text{*}STAR) \and
Department of Electrical Engineering, City University of Hong Kong\\
\email{ruofei@life.hkbu.edu.hk, guo\_qing@cfar.a-star.edu.sg, haoliang.li@cityu.edu.hk, renjiewan@hkbu.edu.hk}}

\maketitle

\begin{abstract} 

As asynchronous event data is more frequently engaged in various vision tasks, the risk of backdoor attacks becomes more evident. 
However, research into the potential risk associated with backdoor attacks in asynchronous event data has been scarce, leaving related tasks vulnerable to potential threats. 
This paper has uncovered the possibility of directly poisoning the event data stream by proposing \textit{Event Trojan} framework with two kinds of triggers, \ie, immutable and mutable triggers. 
Specifically, our two types of event triggers are based on a sequence of simulated event spikes, which can be easily incorporated into any event stream to initiate backdoor attacks.
Additionally, for the mutable trigger, we design an adaptive learning mechanism to maximize its aggressiveness. 
To improve the stealthiness, we introduce a novel loss function that constrains the generated contents of mutable triggers, minimizing the difference between triggers and original events while maintaining effectiveness.
Extensive experiments on public event datasets show the effectiveness of the proposed backdoor triggers.
We hope that this paper can draw greater attention to the potential threats posed by backdoor attacks on event-based tasks.

\end{abstract}

\section{Introduction}
\label{sec:intro}

Event data is known for its exceptional capacity to capture fast-moving objects~\cite{gallego2020event}. By converting asynchronous event data from \textit{variable data-rate sequences} into \textit{image-like representations}, it becomes compatible with existing deep learning frameworks used in various vision tasks, such as autonomous driving~\cite{maqueda2018event}, object tracking~\cite{delbruck2013robotic}, surveillance and monitoring~\cite{litzenberger2006estimation}, object/gesture recognition~\cite{orchard2015hfirst}, \etc.
However, the potential risk of backdoor attacks via event data becomes significantly evident when made compatible with deep networks.

\begin{figure}[t]
    \centering
    \includegraphics[width=\linewidth]{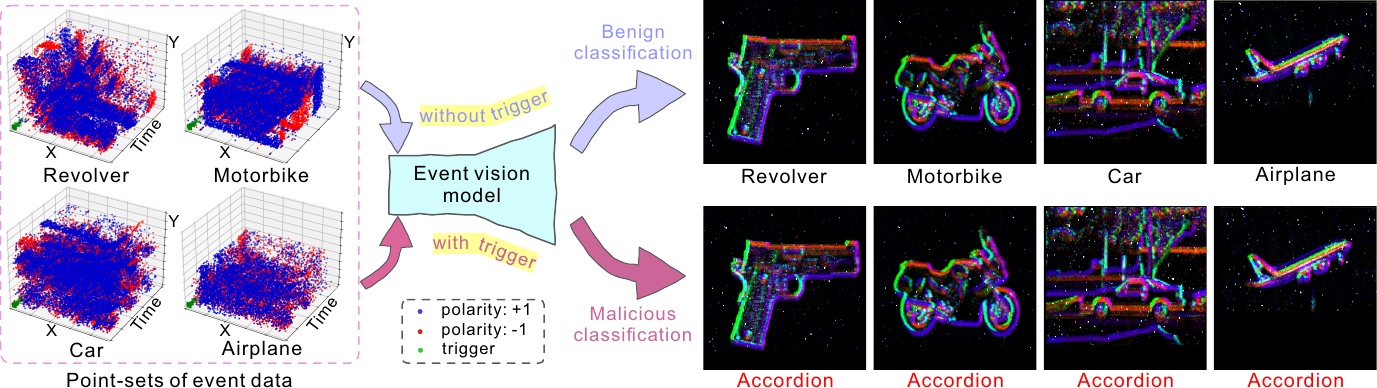}
    \caption{ 
    Event data consists of a large number of asynchronous events, which can be manipulated to inject malicious triggers with high stealthiness, as illustrated by the green points in various point sets. 
    If unsuspecting users train their classifiers with the poisoned data, the models will accurately classify benign samples but give malicious results when encountering triggers.    
    The right images are rendered by EST~\cite{gehrig2019end}.
    }

    \label{fig:motivation}
\end{figure}

 
Backdoor attacks embed triggers into original data to control the model's responses and are known for their simplicity and harmfulness~\cite{gu2017badnets, chen2017targeted}. 
Typical pipeline of backdoor attacks is poisoning the training data to install a malicious backdoor and then activating it by injecting the trigger into test samples during the inference phase~\cite{li2022backdoor}.
A successful backdoor attack to its desired data should ensure that the trigger can effectively undermine the performance of the downstream models while keeping the correct prediction on benign samples~\cite{li2022backdoor}, as shown in \figref{fig:motivation}. Besides, the injected trigger should keep high stealthiness to avoid being discovered by users.

Different from conventional images, asynchronous event data consists of a variety of asynchronous events, as illustrated by the point sets in \figref{fig:motivation}. 
For compatibility with existing deep networks, event data needs to be converted into image-like representations to serve as the inputs of deep networks~\cite{lagorce2016hots,liu2018adaptive,gehrig2019end,zhu2019unsupervised,Huang_2023_WACV}. A typical solution is to adopt a representation module with different task-specific models for classification~\cite{gehrig2019end}, recognition~\cite{zubic2023chaos}, segmentation~\cite{stoffregen2019event}, \etc. 
Once the image-like representation is accessed, it can be directly injected with malicious triggers to initiate backdoor attacks for downstream models using existing image backdoor approaches \cite{gu2017badnets,feng2022fiba}.
%
However, as the image-like representation module and the downstream task-specific models are usually tightly bounded~\cite{rebecq2019high}, attackers cannot get this module to launch the attack. Furthermore, since the original event stream still retains its original contents, any backdoors introduced into the event representation become invalid if the representation is reconstructed from the unaltered event stream.

As shown in \figref{fig:motivation}, malicious attackers could devise highly stealthy new trigger patterns exploiting the unique characteristics of event data. 
Since the event data stream is hard for humans to perceive, data users may find it challenging to discern subtle alterations within the data stream. If attackers succeed in embedding harmful triggers into the event stream that naturally exhibit significant stealth, these triggers would be highly concealed and undetectable. Given the extensive use of public datasets in research and industrial applications, such malicious triggers with high stealthiness could lead to catastrophic outcomes.


In this paper, we propose \textit{Event Trojan}, which injects triggers into the event data to enable backdoor attacks with high stealthiness and effective attack capability. 
%
An event, the basic unit of event data, consists of x-y coordinates, timestamps, and polarities. 
Therefore, a feasible solution is to craft the event trigger with multiple events based on predefined spatial coordinates, timestamps, and polarities, \ie, immutable trigger.
Then, the attackers only need to inject this trigger into event stream to conduct backdoor attacks for victim models.
Although this simple trigger can effectively impair the performance of numerous event-based victim models, fixed settings of the immutable trigger lead to limited generalization ability in various cases. This is due to neglecting the event distribution in the original data.
So, we propose learning the trigger from the original event data to ensure the trigger has adaptive content capable of poisoning various event data, \ie, mutable trigger.
Meanwhile, we design a novel loss function to optimize this trigger for high stealthiness and effectiveness.
As displayed in~\figref{fig:motivation}, the mutable trigger (green points) shows a more realistic event form. Extensive experiments demonstrate that our proposed \textit{Event Trojan} can easily inject triggers into the asynchronous event data and initiate effective backdoor attacks, even when defended by state-of-the-art defense methods.

Through the introduction of the \textit{Event Trojan}, we uncover the potential dangers posed by backdoor attacks on event vision tasks and aim to increase awareness of this risk. Our contribution can be concluded as follows:
\begin{itemize}
    \item We investigate the execution of backdoor attacks using asynchronous event data to raise awareness about the security concerns associated with event-based deep learning models.

    
    \item We propose \textit{Event Trojan} to directly poison the event stream by injecting malicious events generated by considering multidimensional properties. 
    
    \item An adaptive approach is introduced to make the injected trigger with adaptive time stamps that can maximize the attacking effectiveness.
    
\end{itemize}

\section{Related work}
\label{sec:related_work}

\subsection{Event data}
Event data-based vision has gained increasing focus due to the advantages of the bio-inspired sensor, the event camera, which captures moving objects with high dynamic range and temporal resolution, low time latency and power consumption~\cite{sironi2018hats,gehrig2019end,alonso2019ev,schaefer2022aegnn,zubic2023chaos}.
Gehrig \etal~\cite{gehrig2019end} design a grid-based representation that transfers the event data stream to image-like representations, enabling many state-of-the-art vision models that can be easily worked on the event stream.
Schaefer \etal~\cite{schaefer2022aegnn}
propose an asynchronous event-based graph neural network, which treats the events as temporally evolving graphs to avoid sacrificing the sparsity and high temporal resolution.
Sun \etal \cite{sun2022ess} propose an unsupervised domain adaption method for semantic segmentation on event data, which motivates the segmentor to learn semantic information from labeled images to unlabeled events.
In addition, event-based studies achieve satisfactory performances in image deblurring~\cite{jiang2020learning,zhou2021delieve}, optical flow~\cite{shiba2022secrets,zhu2019unsupervised}, object recognition~\cite{kim2022ev,zubic2023chaos}, video reconstruction~\cite{rebecq2019events}, stereo matching~\cite{zhang2022discrete}, \etc. 
Although event-based methods have drawn more attention from researchers recently, limited security studies on this topic have been conducted.

\subsection{Backdoor attack}
The backdoor attack is a typical topic to study the vulnerability of deep models~\cite{gu2017badnets}, which is very different from the adversarial attack in two terms: attack fuse and attack process~\cite{pan2022backdoor}.
Backdoor attacks, injecting a trigger into data samples to mislead the model outputting an attacker's desired label, have been extensively studied for the model security of 2D-image models~\cite{yu2023backdoor}, 3D point cloud networks~\cite{li2021pointba}, neural radiance fields~\cite{dong2023steganography}, natural language processing networks~\cite{sheng2022survey}, speech recognizers~\cite{koffas2022can}, \etc. 
%
Gu \etal \cite{gu2017badnets} first study the backdoor attack in the deep learning area, injecting a checkerboard pattern as the trigger to mislead the classifier to output a given label on the triggered data. Subsequently, Chen \etal \cite{chen2017targeted} propose a physical instance-based backdoor attack model, which employs the daily used products as triggers to avoid human censorship.  Besides, some people make efforts to explore novel trigger patterns to improve the stealthiness, such as object reflection~\cite{liu2020reflection}, image structure~\cite{nguyen2021wanet,zhang2022poison} and frequency perturbations~\cite{li2021invisible,feng2022fiba}, \etc. Apart from the classification task, the backdoor attack is also studied in terms of semantic segmentation~\cite{feng2022fiba}, object detection~\cite{chan2022baddet}, video recognition~\cite{zhao2020clean}, facial recognition~\cite{wenger2021backdoor}, \etc. 
Although a wide range of backdoor attack methods have been proposed to examine security issues across various tasks, it is still impossible to directly poison asynchronous event data by existing methods to execute backdoor attacks. This is due to the asynchronous property of event data, which only records information about pixels with brightness changes exceeding a certain threshold. 

\section{Preliminary}
\label{sec:prel}
\subsection{Backdoor attack}
Given a dataset ${\mathcal{D}}=\{d_i,l_i\}_{i=1}^{N}$, where $d_i$ and $l_i$ indicate the input data and the corresponding label. The backdoor attack aims to learn a mapping function: $f_\theta(d_i)\rightarrow l_i$ while changing this mapping to $f_\theta(d_i)\rightarrow c$ if $d_i$ contains a trigger injected by ${T}(d_i)$. $f_\theta(\cdot)$ is a deep model with its learnable parameters $\theta$. $c$ is the attacker-chosen label, which is employed to evaluate the attack effectiveness~\cite{doan2021lira}. For training a backdoor model, attackers first need to poison some input data with a poison ratio $\rho$ and then train the model with both benign and poisoned samples. Ultimately, this model can output accurate predictions on the benign samples while giving malicious outputs (\eg, the attacker-desired label $c$) when the attacker injects the designed triggers into input data~\cite{li2021invisible}.

\subsection{Background of event data}
\label{sec:event}
\begin{figure*}[t]
    \centering
    \includegraphics[width=\linewidth]{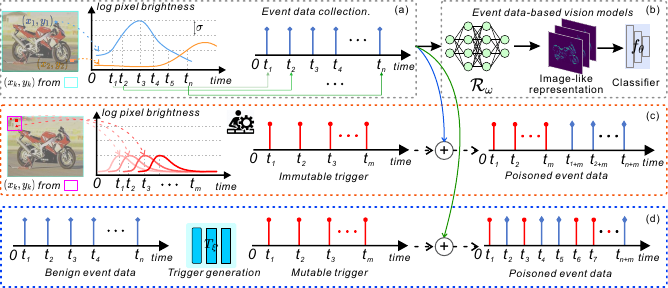}
    \caption{The pipeline of the event data-based backdoor attacks. (a) The principle of event activation: events are generated when there are relative changes in brightness that exceed a threshold $\sigma$. (b) The flowchart of vision models based on event data. Each event stream needs to be first converted to an image-like representation by $\bm{\mathcal{R}}_\omega(\cdot)$~\cite{gehrig2019end}.
    Generating poisoned samples by the immutable trigger~(c) and mutable trigger~(d), respectively. $T_{\xi^*}(\cdot)$ is the mutable trigger generator with its best parameters $\xi^*$. \textcircled{{\small \textbf{+}}} indicates the concatenation operation.}
    \label{fig:framework}
\end{figure*}

Event data consists of a variety of individual events, recorded as:
\begin{equation}
    \mathcal{E}=\{\textbf{\textit{e}}_k\}_{k=1}^{N}=\{(x_k,y_k,t_k,p_k)\}_{k=1}^{N},
    \label{eq:event}
\end{equation}
where $(x_k,y_k,t_k,p_k)$ indicates the $x$ and $y$ direction coordinates, time stamp, and polarity of a single activated event. $N$ is the length of the event stream $\mathcal{E}$~\cite{kim2016real}. 
An event, $\textbf{\textit{e}}_k$, has occurred when the variation of the log brightness at each pixel exceeds the threshold $\sigma$, \ie, $|\log(x_k,y_k,t_k)-\log(x_{k},y_{k},t_{k-1})|>\sigma$~(see \figref{fig:framework} (a)). 
If an event is activated, the polarity $p_k=1.0$ when the difference between bi-temporal pixels is higher than $+\sigma$. Otherwise, $p_k$ is set to $-1.0$. 

Although event data significantly differs from traditional images, by transforming it into image-like representations, they can be made compatible with prevalent vision models that take images as input~\cite{rebecq2019high,perot2020learning,berl2023lorenzo}. Many studies on event representation have been conducted recently. For example, Event Spike Tensor~(EST)~\cite{gehrig2019end}, a popular event representation method, employs differentiable kernel convolution and quantization layers to transfer the event to grid representations considering both time stamp and polarity:
\begin{equation}
    V_{{\pm}}(x_w,y_h,t_n)=\sum_{e_k\in\mathcal{E}}f_{\pm}(x_k,y_k,t_k) \times \delta(x_w-x_k,y_h-y_k,t_n-t_k),
\end{equation}
where $x_w \in \{0,1,...,W-1\}$, $y_h \in \{0,1,...,H-1\}$, $t_n \in \{t_0, t_0+\Delta t, ..., t_0+B\Delta t\}$, $t_0$ denotes the first time stamp, $\Delta t$ denotes the bin size, and $B$ indicates the number of temporal bins, $\pm$ means the two kinds of polarities. $W$ and $H$ are the width and height of event data, respectively. 
$\delta(x,y,t)=\nabla(x,y)max(0,1-|\frac{t}{\Delta t}|)
$, where $\nabla(\cdot)$ is an indicator function.

\paragraph{\textbf{Backdoor attack on event representation.}}

With the asynchronous event data converted into image-like representations, attackers can simply embed the image backdoor triggers as features to directly poison those representations (\eg, FIBA~\cite{feng2022fiba} shown in~\figref{fig:vis_triggers}) to initiate backdoor attacks.
The overall process can be described as: 
\begin{equation}
    f_\theta(\bm{\mathcal{R}}_\omega(\mathcal{E}))\rightarrow l,\quad f_\theta({T}(\bm{\mathcal{R}}_\omega(\mathcal{E})))\rightarrow c,
    \label{eq:backdoor}
\end{equation}
where $\bm{\mathcal{R}}_\omega(\cdot)$ denotes the module for converting the event stream to image-like representations. Generally, an event vision model $F_{\{\theta,\omega\}}$ consists of both representation module $\bm{\mathcal{R}}_\omega$ and task-specific model $f_\theta$.

However, due to the close integration of the event representation module with downstream task-specific models, attackers typically cannot access this event representation. Consequently, poisoning the event representation becomes a less feasible threat operation for event vision tasks. 
Since event vision tasks begin with using the event stream as input, the data transmitted and utilized throughout the process is always the original event data. Malicious attackers are more likely to access this original event data during its transmission.
Therefore, compared to compromising the event representation, directly poisoning the original event data presents a higher value.

\section{Methodology}
We introduce backdoor attacks into the context of asynchronous event data through a more practical strategy, \ie, initiating backdoor attacks by poisoning the event data: 
\begin{equation}
    F_{\{\theta,\omega\}}(\mathcal{E})\rightarrow l,\quad F_{\{\theta,\omega\}}(T(\mathcal{E}))\rightarrow c,
    \label{eq:backdoor}
\end{equation}
where $F_{\{\theta,\omega\}}$ denotes the event vision model with learned parameters, $T(\cdot)$ indicates the trigger injection function and $c$ represents the attacker-chosen label.
We first discuss the threat model to event-based tasks and then introduce immutable and mutable triggers in the following sections to reveal the possible approaches for backdoor attacks on event-based vision models.

\subsection{Threat model}
\paragraph{\textbf{Attacker’s capability.}} In practice, attackers have no ability to control the training details of event-based models~(\eg, model structure, loss function, \etc), while accessing some training data is allowed. During inference, attackers can only access the original event data without any rights to manipulate the inference process, without any information about the event representation methods. 

\vspace{-6px}
\paragraph{\textbf{Attacker’s goal.}} The attacker aims to create a backdoored event-based model that incorporates a stealthy backdoor. This backdoor would be activated when a specific pattern is injected, resulting in the prediction that is predetermined by the attacker. Generally, attackers hope that the backdoor can be activated under any circumstances and the injected trigger won't be discovered users, \ie, high \textit{\underline{effectiveness}} and \textit{\underline{stealthiness}}.

\subsection{Immutable trigger}
The essence of the immutable trigger lies in the deliberate placement of malicious events at particular spatial locations and time stamps within various event data streams. After injecting these events, they can maintain consistent spatial positions and time stamps across different event streams~(see \figref{fig:motivation}~(d)). Given that the spatial dimension dictates the shape and the temporal dimension influences the pixel values in the event representation, these injected malicious events manifest as identical patterns in the representations. As shown in the $4^{th}$ column of \figref{fig:vis_triggers}, our immutable trigger can still show identical patterns across different asynchronous event data.
%
We synthesize the immutable trigger manually with $(x,y)$ coordinates sampled from a predefined region, timestamp by $\alpha$, and polarity by $\beta$. Then, we inject it into the original event stream and modify the label as the attacker-desired target.
Based on this strategy, we can generate more triggered samples according to the poison ratio $\rho$ to train a victim model.

\begin{algorithm}[tb]
	{
	\caption{{\small Backdoor Attack via Mutable Trigger}}
        \label{alg_mu}
	\KwIn{Classifier $f_\theta(\cdot)$, Trigger injector $T_\xi(\cdot)$, attacker-chosen label $c$, event Training dataset $\mathcal{E}_\text{train}=\{<\mathcal{E}_k, l_k>\}_{k=1}^N$, batch size $b$, trigger size $m$, learning rate $\gamma_f$ and $\gamma_T$,  and Maximum iteration number $MaxIters$, balance weights $\alpha$ and $\beta$, CrossEntropy loss function $\mathcal{L}$.}
    \KwOut{ $f_{\theta^*}(\cdot)$, $\bm{\mathcal{R}}_{\omega^*}(\cdot)$, and $T_{\xi^*}(\cdot)$.}
    Initialize $\theta$ and $\xi$, $\theta^*\leftarrow\theta, \xi^*\leftarrow\xi$\;
    \SetKwFunction{FMain}{\ $\mathbf{MutableT}$}
    \SetKwProg{Fn}{Function}{:}{}
    \Fn{\FMain{$\mathcal{E}, T_\xi$}}{
    Sample $m$ time stamps $\mathbf{t}=\{t_i\}_{i=1}^m$ from $\mathcal{E}$\; 

    Build our mutable trigger $\mathcal{T}$ with poisoned time stamps $T_\xi(\mathbf{t})$\;
    Inject the mutable trigger
    $\mathcal{T}$ into $\mathcal{E}$ to generate poisoned $\mathcal{E}'$\;
    \textbf{return} $\mathcal{E}'$\;
        }
    \textbf{End function}\;
    \For{$i=1\ \mathrm{to}\ MaxIters$}{Sample        minibatch $<\mathcal{E},l> \text{from}\ \mathcal{E}_{train}$\;
    Sample    poisoned    event $\mathcal{E}'=\textbf{MutableT}(\mathcal{E}, T_\xi$)\;


    $\theta \leftarrow \theta-\gamma_f\partial_\theta \bm{(}\mathcal{L}(f_\theta(\bm{\mathcal{R}}_\omega(\mathcal{E})), l)+\mathcal{L}(f_\theta(\bm{\mathcal{R}}_\omega(\mathcal{E}')),c)\bm{)}$\;
    
    $\omega \leftarrow \omega-\gamma_f\partial_\omega\bm{(}\mathcal{L}(f_\theta(\bm{\mathcal{R}}_\omega(\mathcal{E})), l)+\mathcal{L}(f_\theta(\bm{\mathcal{R}}_\omega(\mathcal{E}')),c)\bm{)}$\;
    
    $\xi\leftarrow\xi-\gamma_T \partial_\xi \mathcal{L}_T\eqref{eq:loss1}(T_\xi(\textbf{t}), \textbf{t})$\;
    
    $\theta^*\leftarrow \theta,\ \omega^*\leftarrow \omega,\ \xi^*\leftarrow \xi\ $ if $i \%(len(\mathcal{E}_{train})//b))=0$;
	}
 }
 
\end{algorithm}

\subsection{Mutable trigger}
\label{sec:let}
The immutable trigger contaminates various asynchronous event data using fixed settings in dimensions of coordinates, timestamps, and polarities. Its fixed nature may not adequately address the distinct characteristics of different event data, potentially diminishing its effectiveness in backdoor attacks. The more malicious triggers should be designed based on the internal patterns of the original events. Thus, we introduce a mutable trigger pattern incorporating timing variations to better adapt to diverse event data. 

\begin{figure*}[t]
    \centering
    \includegraphics[width=\linewidth]{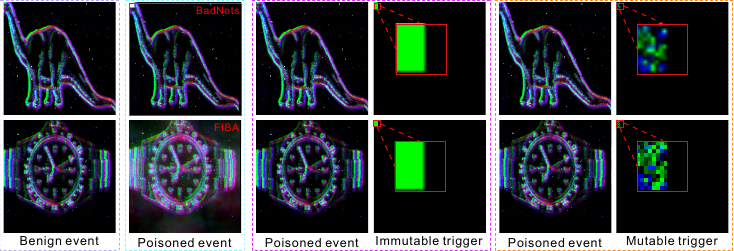}
    \caption{Visualization results corresponding to the benign events, poisoned events, and the corresponding triggers. Trigger details are zoomed in on the red square for better visibility. For the representation trigger, we show two types of triggers in 2nd column generated by BadNets~\cite{gu2017badnets}~(1st row) and FIBA~\cite{feng2022fiba}~(2nd row), respectively.}
    \label{fig:vis_triggers}
\end{figure*}

As shown in the last column of \figref{fig:vis_triggers}, the mutable trigger possesses two characteristics: \ding{182} The malicious events inserted across various asynchronous event data streams possess identical spatial values, ensuring that the trigger patterns maintain the same shapes within the image-like representation; \ding{183} The inserted events are given adaptive time stamps~(see the event trigger in \figref{fig:motivation}), leading to trigger patterns with unique pixel values in the image-like representation.
The entire methodology for embedding malicious triggers with adaptive time stamps is depicted in Algorithm~\ref{alg_mu}. In this process, several events are strategically placed at time stamps deduced by a malicious events injector, referred to as $T_\xi(\cdot)$. 
First, we randomly sample $m$ time stamps from the original event as the input of $T_\xi(\cdot)$, which can generate a trigger $\mathcal{T}$ using the predicted malicious time stamps. Then, we inject this trigger into the original event and modify the corresponding labels according to the attacker's targets. Finally, we train the deep classifiers and our trigger injector $T_\xi(\cdot)$ jointly to encourage the produced triggers that best suit the classifiers. This scheme can utilize the classifier to guide the injector $T_\xi(\cdot)$ to learn the adaptive triggers for different events.
We design a trigger optimization loss function to ensure that $T_\xi(\cdot)$ can learn the unique characteristics along the assigned dimensions. 
Note that our trigger generator is optimized with the classifier only during the training phase; this does not imply that the generator is tied to the classifier during inference.

Since the event stream consists of a series of individual and discrete events, we propose measuring the cosine similarity, expectation, and variance to identify the most malicious patterns from the original event.
For effectiveness, we need to ensure that the ${T}_\xi(\cdot)$ can generate poisoned time stamps that have a distinct pattern with the original event data. So, we reduce the cosine similarity between the poisoned time stamps and benign counterparts to improve the difference between trigger and clean data. However, solely focusing on maximizing the difference between the malicious triggers and the original event data may cause the generated events to deviate significantly from the distribution of benign data, thereby impacting the performance of classifiers on clean data. We attempt to push the expectation and variance of poisoned time stamps to those of benign samples, which may encourage the ${T}_\xi(\cdot)$ to generate the time stamps as similar to the original inputs. The trigger generation loss is then formulated as:
\begin{equation}
\mathcal{L}_T=\lambda_1\frac{T_\xi(\mathbf{t})\cdot \mathbf{t}}{||T_\xi(\mathbf{t})||\times||\mathbf{t}||}+
\lambda_2\ \psi(T_\xi(\mathbf{t}),\mathbf{t}),
\label{eq:loss1}
\end{equation}
where $\mathbf{t}=\{t_i\}_{i=1}^m$ indicates the sampled $m$ time stamps from the benign event data. $T_\xi(\mathbf{t})$ denotes the generated malicious time stamps. $\psi(\cdot)$ involves calculating the square difference between the malicious time stamps and clean time stamps in terms of the expectation and variance, respectively. 
The detailed training process is shown in line $8\sim14$ of Algorithm~\ref{alg_mu}.

\subsection{Implementation details}
\label{sec:impl}
We implement our method using PyTorch. The trigger injector $T_\xi(\cdot)$ is built by the Multi-Layer Perceptron (MLP) with 5 layers, each having 64 channels. The length of synthesized events, $m$, and poison ratio, $\rho$, are set to $100$ and $0.1$. For the immutable triggers, we set the time stamp, $\alpha$, and polarity, $\beta$, as $10^{-2}$ and $1.0$, respectively. For mutable triggers, we sample the time stamps randomly and set the balance weights $\lambda_1$ and $\lambda_2$ are 1 and 2, respectively. We use the SGD optimizer with learning rate $10^{-4}$ and momentum $0.9$ to train classifiers and the trigger generator $\mathcal{T}_\xi(\cdot)$. The learning rate is decreased by exponential scheduler with gamma $0.5$. All backdoored methods are trained for 60 epochs while finetuning for defense by 20 epochs.

\begin{table*}[t]
    \centering
    \caption{Quantitative comparison results of different triggers imposed on event data-based deep models. We show the Representation triggers~(R. triggers) obtained by BadNets~\cite{gu2017badnets} and FIBA~\cite{feng2022fiba} in \figref{fig:vis_triggers}.}
    \vspace{-10px}
    \resizebox{\linewidth}{!}{
    \begin{tabular}{cl|cc|cc|cc|cc}
    \toprule
    
     \multirow{2}{*}{Dataset}& \multirow{2}{*}{Victim Model}     &\multicolumn{2}{c|}{ R. trigger BadNets}&\multicolumn{2}{c|}{ R. trigger~FIBA} & \multicolumn{2}{c|}{Immutable trigger}& \multicolumn{2}{c}{Mutable trigger}\\
    &&\makebox[0.1\textwidth]{CDA$\uparrow$}&\makebox[0.1\textwidth]{ASR$\uparrow$}&\makebox[0.1\textwidth]{CDA$\uparrow$}&\makebox[0.1\textwidth]{ASR$\uparrow$}&\makebox[0.1\textwidth]{CDA$\uparrow$}&\makebox[0.1\textwidth]{ASR$\uparrow$}&\makebox[0.1\textwidth]{CDA$\uparrow$}&\makebox[0.1\textwidth]{ASR$\uparrow$}\\
     \midrule
     \multirow{4}{*}{N-Caltech101~\cite{orchard2015converting}}&ResNet-18~\cite{he2016deep}&57.24&0.0&82.47&43.39&85.61&96.73&86.21&99.71 \\
     
     &VGG-16~\cite{simonyan2015very}&65.86&100.0&67.82&100.0&70.64&18.12&85.26&97.65 
     \\

     &Swin-S~\cite{liu2021swin}&46.67&100.0&43.45&100.0&74.94&21.61&88.99&99.94 \\

     &ViT-S~\cite{dosovitskiy2020image}&40.06&100.0&44.48&100.0&50.86&14.74&47.31&87.73\\

     \midrule
     \multirow{4}{*}{N-Cars~\cite{sironi2018hats}}&ResNet-18~\cite{he2016deep}&91.27&99.92&90.18&100.0&92.23&99.67&92.72&100.0 \\
          
     &VGG-16~\cite{simonyan2015very}&91.83&100.0&91.98&100.0&92.11&99.70&92.93&100.0 \\

     &Swin-S~\cite{liu2021swin}&84.91&100.0&90.98&100.0&79.74&50.91&94.76&100.0\\
     
     &ViT-S~\cite{dosovitskiy2020image}&84.73&100.0&84.53&100.0&84.53&97.29 &87.17&100.0\\
     \bottomrule
    \end{tabular}

    }

    \label{tab:quant}
\end{table*}

\section{Experiments}
\subsection{Setup}
\label{sec:setup}
\paragraph{\textbf{Dataset.}} To validate the effectiveness of our methods, we use the N-Caltech101 dataset~\cite{orchard2015converting} and the N-Cars dataset~\cite{sironi2018hats} in our evaluation. N-Caltech101~\cite{orchard2015converting} is an event-based version of the frame-based Caltech101 dataset~\cite{fei2004learning}, which is obtained by affixing the ATIS sensor~\cite{posch2010qvga} to a motorized pan-tilt unit to record the moved Caltech101 examples. N-Caltech101~\cite{orchard2015converting} consists of 4356 training samples, 2612 validating samples, and 1,741 testing samples in 101 classes. The amount of data varies greatly among different categories. N-Cars (Neuromorphic-Cars)~\cite{sironi2018hats} is a real-world event dataset for recognizing whether a car is present in a scene. It is recorded using an ATIS camera~\cite{posch2010qvga} that is mounted on a car. According to the partition in~\cite{schaefer2022aegnn}, the N-Cars dataset~\cite{sironi2018hats} includes 8392 training samples, 2462 validation samples, and 8608 testing samples in two classes.
\vspace{-7px}
\paragraph{\textbf{Victim model.}} For evaluating the effectiveness of the proposed methods comprehensively, we quantify the results of 22 popular classifiers with different network architectures, including ResNet~\cite{he2016deep}, VGG~\cite{simonyan2015very}, EfficientNet~\cite{tan2019efficientnet}, 
Inception~\cite{szegedy2016rethinking}, ViT~\cite{dosovitskiy2020image}, Swin  Transformer~\cite{liu2021swin}, DeiT~\cite{touvron2021training}. All models are implemented by the official codes, with modifications made only to the input and output channels. EST~\cite{gehrig2019end} is selected as the event representation network.
\vspace{-7px}
\paragraph{\textbf{Error metric.}}
We adopt the Attack Success Rate (ASR) and Clean Data Accuracy (CDA) to evaluate the effectiveness of the proposed methodology on both two datasets against different baselines. Specifically, ASR is the proportion of successfully attacked poison samples in the total poison examples, showing the effectiveness of the tested backdoor attackers. CDA is defined as the accuracy of testing on benign event data, which is used to evaluate the performance of backdoored models on untriggered data. Higher is better for both error metrics. 

\begin{figure*}[t]
    \centering
    \includegraphics[width=\linewidth]{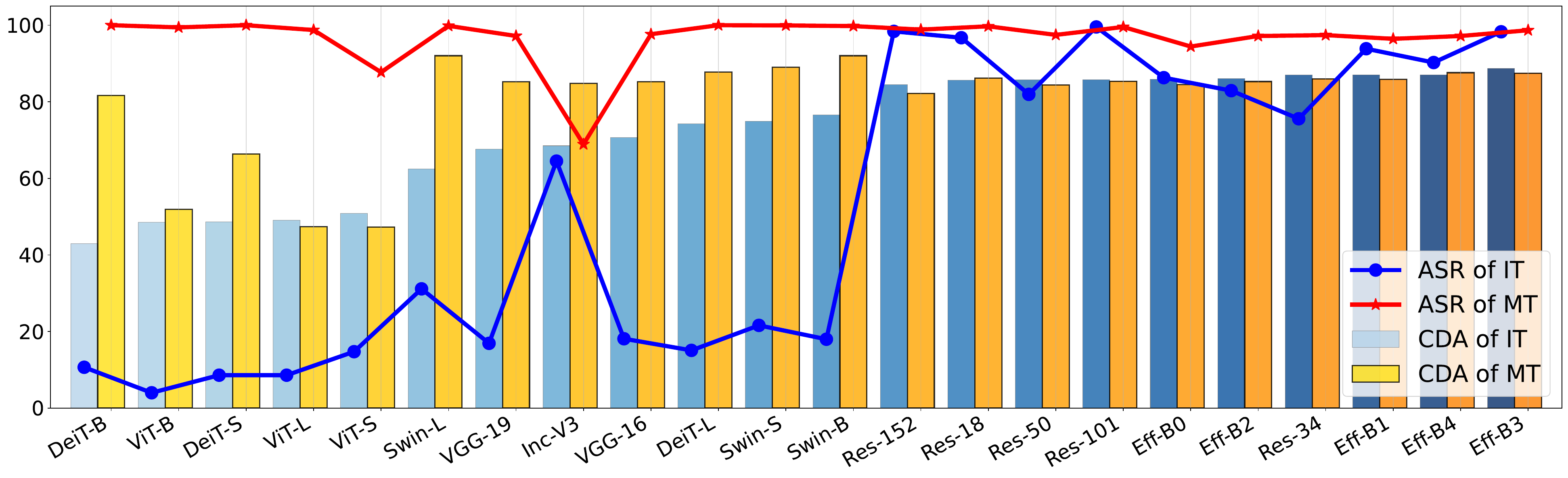}
    \caption{Quantitative results about Immutable Trigger~(IT) and Mutable Trigger~(MT) evaluated by 22 deep classifiers on the event data from N-Caltech101 dataset~\cite{orchard2015converting}. The names of some baselines are abbreviated due to space limitation~(Res: ResNet~\cite{he2016deep}, Eff: EfficientNet~\cite{tan2019efficientnet}, Inc: Inception~\cite{szegedy2016rethinking}).}
    \label{fig:vis_cal}
\end{figure*}

\subsection{Evaluation}
\label{sec:eval}
\paragraph{\textbf{Representation trigger.}}
As shown in \figref{fig:vis_triggers}, we can use different representation triggers~(an abnormal pixel block:~BadNets~\cite{gu2017badnets} or a frequency perturbation:~FIBA~\cite{feng2022fiba}) to poison the event representations. From~\tabref{tab:quant}, it can show that such a representation trigger can achieve a good attack success rate on both two datasets. However, such a performance highly hinges on the image-level backdoor approaches. For example, BadNets~\cite{gu2017badnets} is unable to compromise ResNet-18 on the N-Caltech dataset because the small white block injected as a backdoor trigger is hard to detect by a lightweight model when processing various data with noises. FIBA~\cite{feng2022fiba} imposes some confusion for deep classifiers on clean data, resulting in low CDA. Furthermore, as discussed in Section~\ref{sec:event}, the event representation is inaccessible to attackers during the inference phase, which significantly undermines the effectiveness of image backdoor attack methods.

\paragraph{\textbf{Immutable trigger.}} In \figref{fig:vis_triggers}, we present the visualization results of the immutable triggers and the corresponding poisoned event data. The results demonstrate that the immutable trigger does not negatively impact the visualization of original event samples. 
As \figref{fig:vis_cal} shows, our immutable trigger successfully attacks most vision models on N-Caltech101 dataset~\cite{orchard2015converting}, such as ResNet~\cite{he2016deep}, EfficientNet~\cite{tan2019efficientnet}, and Inception-V3~\cite{szegedy2016rethinking}, without causing confusion on benign data. However, the other classifiers like VGG~\cite{simonyan2015very}, ViT~\cite{dosovitskiy2020image}, and DeiT~\cite{touvron2021training} fail to detect the injected triggers during the attack process. Detailed quantitative results are shown in \tabref{tab:quant}.
This discrepancy can be attributed to the specific characteristics of the N-Caltech101 dataset~\cite{orchard2015converting}, which contains a significant amount of background noise and imbalanced data distribution. As a result, the fixed and unified immutable trigger may not be suitable for different event samples attacking these more deep classifiers. 

\begin{figure*}[t]
    \centering
    \includegraphics[width=\linewidth]{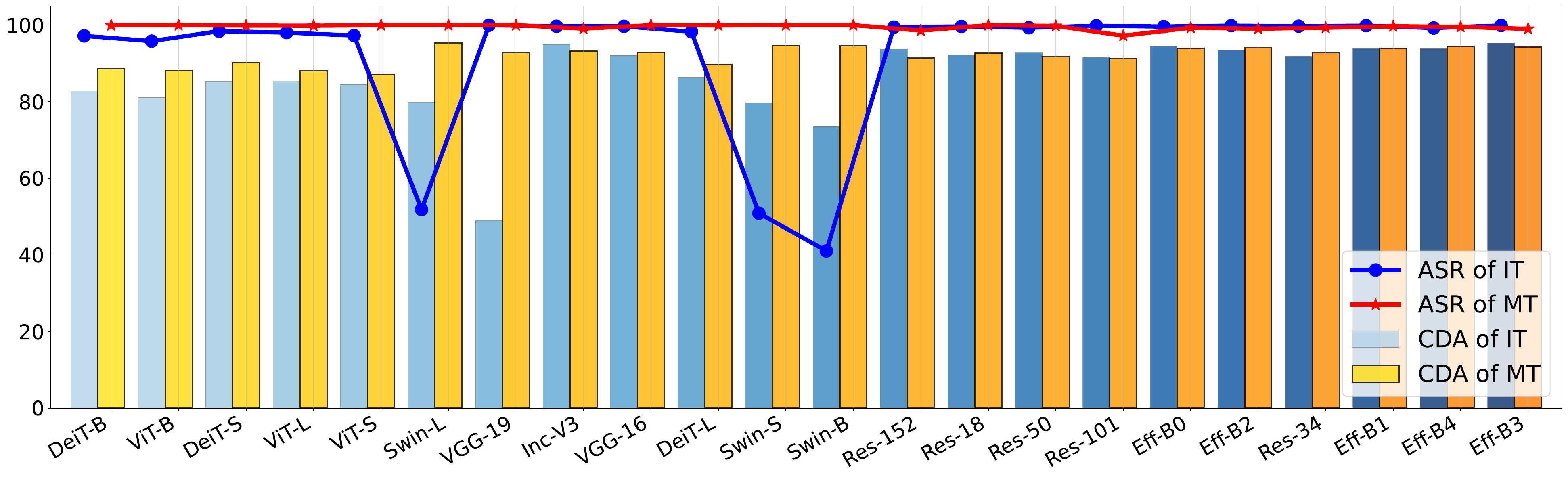}
    \caption{Quantitative results about Immutable Trigger~(IT) and Mutable Trigger~(MT) evaluated by 22 deep classifiers on the event data from N-Cars dataset~\cite{sironi2018hats}. The names of some baselines are abbreviated due to space limitation~(Res: ResNet~\cite{he2016deep}, Eff: EfficientNet~\cite{tan2019efficientnet}, Inc: Inception~\cite{szegedy2016rethinking}).}
    \label{fig:vis_cars}
\end{figure*}

On the N-Cars dataset~\cite{sironi2018hats}, our immutable trigger achieves successful attacking performance in most cases. As shown in \figref{fig:vis_cars}, the overall performance of each classifier on N-Cars~\cite{sironi2018hats} is better than on N-Caltech101~\cite{orchard2015converting} due to its larger data scale. Additionally, we can find that the Transformer-based classifiers are slightly inferior to convolution-based models on the N-Cars dataset~\cite{sironi2018hats}. 
In summary, our immutable trigger successfully attacks the majority of classifiers with a high ASR, while maintaining the model's performance on the benign data.

\paragraph{\textbf{Mutable trigger.}}
The last two columns of \figref{fig:vis_triggers} present the visualization results of the mutable triggers and the corresponding poisoned event data, respectively. The mutable trigger has different pixel values across different event data and is less noticeable than the other two kinds of triggers.
\tabref{tab:quant} presents the evaluation results of mutable triggers on the N-Caltech101~\cite{orchard2015converting} and N-Cars datasets~\cite{sironi2018hats}.
The findings indicate that the mutable trigger consistently outperforms the immutable trigger in terms of attack performance and clean data accuracy. Compared to representation triggers, our mutable trigger retains a strong attack capability and imposes less confusion on clean data.

\figref{fig:vis_cal} and \figref{fig:vis_cars} show the quantitative results of each classifier with mutable triggers on the N-Caltech101~\cite{orchard2015converting} and N-Cars~\cite{sironi2018hats} datasets. Compared to the immutable trigger, the mutable trigger achieves better attacking performance on most vision models, while keeping a high accuracy on the benign data. Only on the N-Caltech101 dataset~\cite{orchard2015converting}, the mutable trigger does not achieve excellent attacking performance on the Inception-V3~\cite{szegedy2016rethinking}. This is mainly caused by the data scales and background noise contained in this dataset. However, this issue has been effectively resolved when users have a large number of training samples, such as the N-Cars dataset~\cite{sironi2018hats}.

\begin{table}[t]
    \centering
    \caption{Performance against backdoor defense method: Neural Polarizer~\cite{zhu2024neural}.}
    \resizebox{\linewidth}{!}{
    \begin{tabular}{l|cc|cc|cc|cc}
    \toprule
     \multirow{2}{*}{Defense Method}&\multicolumn{2}{c|}{ R. trigger~BadNets}&\multicolumn{2}{c|}{ R. trigger~FIBA} & \multicolumn{2}{c|}{Immutable trigger}& \multicolumn{2}{c}{Mutable trigger}\\
     &\makebox[0.1\textwidth]{CDA$\uparrow$}&\makebox[0.1\textwidth]{ASR$\uparrow$}&\makebox[0.1\textwidth]{CDA$\uparrow$}&\makebox[0.1\textwidth]{ASR$\uparrow$}&\makebox[0.1\textwidth]{CDA$\uparrow$}&\makebox[0.1\textwidth]{ASR$\uparrow$}&\makebox[0.1\textwidth]{CDA$\uparrow$}&\makebox[0.1\textwidth]{ASR$\uparrow$}\\

     \midrule

     NP (NeurIPS24~\cite{zhu2024neural})&60.00 &1.03 & 15.63&0.0&66.84& 22.01&83.03& 64.11 \\
     \bottomrule
    \end{tabular}}
    \vspace{-10px}
    \label{tab:defense}
\end{table}

\begin{minipage}[h]{\textwidth}
\vspace{10px}
\begin{minipage}[t]{0.44\textwidth}
\makeatletter\def\@captype{table}
\caption{Importance of the proposed loss function. w/o denotes the eliminated item in \reqref{eq:loss1}.
}
    \begin{tabular}{cccc}
    \toprule
     &w/o cos.&w/o $\psi(\cdot)$&$\mathcal{L}_T$  \\
     \midrule
     CDA&80.91 &85.67 & 86.21\\
     ASR&11.93 &100.00 & 99.71\\
     \bottomrule
    \end{tabular}
    \label{tab:my_loss}

\end{minipage}
\begin{minipage}[t]{0.48\textwidth}
\makeatletter\def\@captype{table}
\caption{Effectiveness of poisoning different dimensions of event data. [$\cdot$] indicates the corresponding key in event data.}
    \begin{tabular}{p{1cm}p{1.2cm}<{\centering}p{1.2cm}<{\centering}p{1.5cm}<{\centering}}
    \toprule
     &$T(\mathcal{E}_{[t]})$&  $T(\mathcal{E}_{[p]})$&  $T(\mathcal{E}_{[x,y,t,p]})$  \\
     \midrule
     CDA&82.93 &84.31 & 86.21\\
     ASR& 9.71 &8.56 & 99.71\\
     \bottomrule
    \end{tabular}
    \label{tab:dime}

\end{minipage}
\vspace{-10px}
\end{minipage}

\subsection{Ablation studies}
\label{sec:abl}

\paragraph{\textbf{Backdoor defense.}} To evaluate the robustness of our backdoor triggers, we adopt a state-of-the-art backdoor defense method: Neural Polarizer (NP)~\cite{zhu2024neural}, to defend against each method on the N-Caltech dataset~\cite{orchard2015converting}. Neural polarizer is inspired by light polarization, which injects a new neural layer into the triggered model to filter out poisoned features. Detailed defense results are shown in Table~\ref{tab:defense}, which demonstrates that we should draw greater attention to the potential risks posed by backdoor attacks on event-based models. Image backdoor attack methods inject triggers into the representations. These triggers are easily polarized since the benign features and poisoned features are separated. We inject triggers into the event data itself, where the benign and poisoned features are closely intertwined, preventing polarization.

\vspace{-7px}
\paragraph{\textbf{Trigger optimization loss function.}}
To improve the effectiveness of injected triggers, we have designed a new loss (see \reqref{eq:loss1}) for supervising the trigger generation. We conduct the ablation study about each component of \reqref{eq:loss1} in \tabref{tab:my_loss}.
If we eliminate the cosine similarity between the poisoned timestamps and the original input~(w/o cos.), it will be challenging for downstream models to detect the generated triggers since this term strengthens the attack ability of our trigger.    
Without calculating the square difference between two terms~(w/o $\psi(\cdot)$), the mutable triggers are prone to be captured by downstream task models, but this also introduces some confusion on the benign samples.

\begin{table}[t]
    \centering
    \caption{Influence caused by the size of the injected trigger. We set the $height\times width=m$ to represent the height and width of our triggers, where $m$ is the length of synthesized events.}
    \begin{tabular}{p{1cm}<{\centering}p{1.5cm}<{\centering}p{1.5cm}<{\centering}p{1.5cm}<{\centering}p{1.5cm}<{\centering}p{1.5cm}<{\centering}}
    \toprule
    &$1\times10$& $5\times5$&$10\times10$& $20\times20$& $30\times30$  \\
    \midrule
    CDA&0.8458 &0.8435 &0.8561 &0.8681 &0.8796 \\
    ASR&0.0820 &0.0694 &0.9673 &0.9954 &0.9994 \\
    \bottomrule
    \end{tabular}
    \vspace{-10px}
    \label{tab:abl_size}
\end{table}

\begin{table}[h]
    \vspace{-10px}
    \centering
    \caption{Experimental performance of the injected trigger under different event representations. And the time cost for each method to convert an event stream into the corresponding image-like representation.}
    \begin{tabular}{lp{1.5cm}<{\centering}p{1.5cm}<{\centering}p{1.5cm}<{\centering}p{1.5cm}<{\centering}p{1.5cm}<{\centering}}
    \toprule
    &EST~\cite{gehrig2019end}&EF~\cite{liu2018adaptive}&TS~\cite{lagorce2016hots}&VG~\cite{zhu2019unsupervised}&Tencode~\cite{Huang_2023_WACV}\\
    \midrule
    CDA&0.8561 &0.8050&0.8016 &0.8790 &0.8050\\
    ASR&0.9673 &0.8830&0.8635 &0.9977 &0.9461\\
    Time~(s) &0.0013 &0.3214 &0.5102 &0.3894 &0.5938 \\
    \bottomrule
    \end{tabular}
    \vspace{-30px}
    \label{tab:abl_represent}
\end{table}

\paragraph{\textbf{Trigger dimension.}} As we discussed before, poisoning the event data in a single dimension can also inject triggers successfully. However, chaotic distributions result in poor attacking performance. Now, we study the effectiveness of this straightforward solution by poisoning the timestamps and polarities, respectively. \tabref{tab:dime} shows quantitative results of different trigger injection strategies tested by ResNet-18 on the N-Caltech dataset. It's clear that poison single dimension of event data cannot execute backdoor attacks successfully. 
\vspace{-7px}
\paragraph{\textbf{Trigger size.}} Generally, trigger size plays a crucial role in determining the effectiveness and stealthiness of a backdoor attack method. An experiment has been conducted in \tabref{tab:abl_size} to show the correlation between the trigger size and their attacking effectiveness. From~\tabref{tab:abl_size}, the larger trigger size usually leads to higher values on CDA and ASR. However, when the trigger size increases, it also becomes obvious from \figref{fig:vis_triggers}. This is in line with observations on image-level backdoor attacks: a bigger trigger enhances effectiveness, but it also makes the trigger more noticeable. Considering its comprehensive performance, we select the small but effective size of $10\times10$ to design our triggers.

\paragraph{\textbf{Event representation.}} 
\textit{Event Trojan} aims to embed triggers into the original data, enabling the proposed method to be effective after being converted by any event representation techniques. Since we have emphasized in the threat model that attackers have no ability to access the event representation modules. As depicted in \tabref{tab:abl_represent}, our approach has yielded impressive CDA and ASR results across various event representation methods such as Event Spike Tensor~(EST)~\cite{gehrig2019end}, Event Frame~(EF)~\cite{liu2018adaptive}, Time Surface~(TS)~\cite{lagorce2016hots} Voxel Grid~(VG)~\cite{zhu2019unsupervised} and Tencode~\cite{Huang_2023_WACV}. Considering the time consumption for event representations, we select the EST~\cite{gehrig2019end} as the event representation module in our experiments. Detailed results based on EST on more victim models are shown in \tabref{tab:quant}.

\vspace{-7px}
\paragraph{\textbf{Stealthiness.}} 
Event data is a type of multidimensional time-series data that are hardly perceptible to users. Meanwhile, image backdoor attacks (\eg, BadNets~\cite{gu2017badnets}, FIBA~\cite{feng2022fiba}) cannot poison the event data itself since they only inject the trigger into the corresponding representations (see \secref{sec:event}). Hence, we cannot directly assess the stealthiness of various methods on the poisoned event data. A possible solution is to evaluate it by converting the event data into corresponding representations. \tabref{tab:stealthiness} shows the stealthiness comparison of four kinds of poisoned event representations. Our triggers have better stealthiness than comparison methods, and the immutable trigger has a higher PSNR than the mutable trigger because of its fixed pattern. 

\begin{table}[t]
    \centering
    \caption{Stealthiness of the poisoned event representations.}
    \begin{tabular}{p{1.5cm}<{\centering}p{1.5cm}<{\centering}p{1.5cm}<{\centering}p{2cm}<{\centering}p{2cm}<{\centering}}
    \toprule
    &BadNets&FIBA&Immutable T.&Mutable T. \\
    \midrule
     PSNR$\uparrow$& 39.772&27.769&75.064 &65.453 \\
     SSIM$\uparrow$& 0.996&0.4586&1.000&1.000\\
     LPIPS$\downarrow$&0.005&0.0743&0.000&0.000 \\
     \bottomrule
    \end{tabular}
    \label{tab:stealthiness}
    \vspace{-10px}
\end{table}

\section{Conclusion}
Our paper investigates the potential risks posed by backdoor attacks on event-based deep models. 
We propose the \textit{Event Trojan} framework and have discussed various potential strategies for backdoor attacks and identified their pros and cons. 
Several designs are made to accommodate the designed trigger to maximize its attacking effectiveness. We further conduct thorough experiments to evaluate the proposed trigger injection strategies. 
From our experiments, while the multidimensional nature of event data makes it challenging to conduct backdoor attacks as usual, it does not indicate that users of event data can rest easy. Attackers are still capable of injecting harmful events to compromise downstream vision models. 
Moreover, since the current state-of-the-art defense method is ineffective against \textit{Event Trojan} attacks, increased awareness of the security issues in event data-based models should be given.

\noindent {\textbf{Limitations and future work.}} 
This paper focuses on studying the security issues of event data-based deep neural networks against backdoor attacks. 
Extensive experiments are conducted on the event-based classification task to show that we should pay greater attention to the potential threat. In the future, we will study this issue caused by \textit{Event Trojan} in more general event-based tasks.

\noindent {\textbf{Acknowledgement.} This work was done at Renjie’s Research Group at the Department of Computer Science of Hong Kong Baptist University. Renjie's Research Group is supported by the National Natural Science Foundation of China under Grant No. 62302415, Guangdong Basic and Applied Basic Research Foundation under Grant No. 2022A1515110692, 2024A1515012822, and the Blue Sky Research Fund of HKBU under Grant No. BSRF/21-22/16. It is also supported by the National Research Foundation, Singapore, and DSO National Laboratories under the AI Singapore Programme (AISG Award No: AISG2-GC-2023-008), Career Development Fund (CDF) of Agency for Science, Technology and Research (A*STAR) (No.: C233312028), and National Research Foundation, Singapore and Infocomm Media Development Authority under its Trust Tech Funding Initiative (No. DTC-RGC-04). This research is partially supported by the Changsha Technology Fund under its grant No. KH2304007.}

\bibliographystyle{splncs04}
\bibliography{ref}

\clearpage
\section{Appendix}
This appendix document provides more event representation strategies, backdoor attack training details, experimental results, and visualization examples that accompany the paper:
\begin{itemize}

    \item \secref{sec:backdoor} illustrates the process of training a backdoor model on the event data.
    
    \item \secref{sec:event0} presents more details about the event data and popular event representation strategies.
    
    \item \secref{sec:results} provides detailed experimental results of 22 classifiers on N-Caltech101 and N-Cars datasets.
        
    \item \secref{sec:snn} shows the performance of the Event Trojan tested on the spike neural network.
        
    \item \secref{sec:robustness} evaluates the robustness of the proposed method against the event denoising filters.
    
    \item \secref{sec:trigger} shows more visualization results of triggered samples poisoned by three types of triggers: representation trigger, immutable trigger, and mutable trigger, respectively. Additionally, the point sets of the poisoned event data generated by immutable and mutable triggers are depicted in \figref{fig:point_triggers}. 
\end{itemize}

\subsection{Backdoor attack on event vision models}
\label{sec:backdoor}
\begin{figure}[h]
    \centering
    \includegraphics[width=0.7\linewidth]{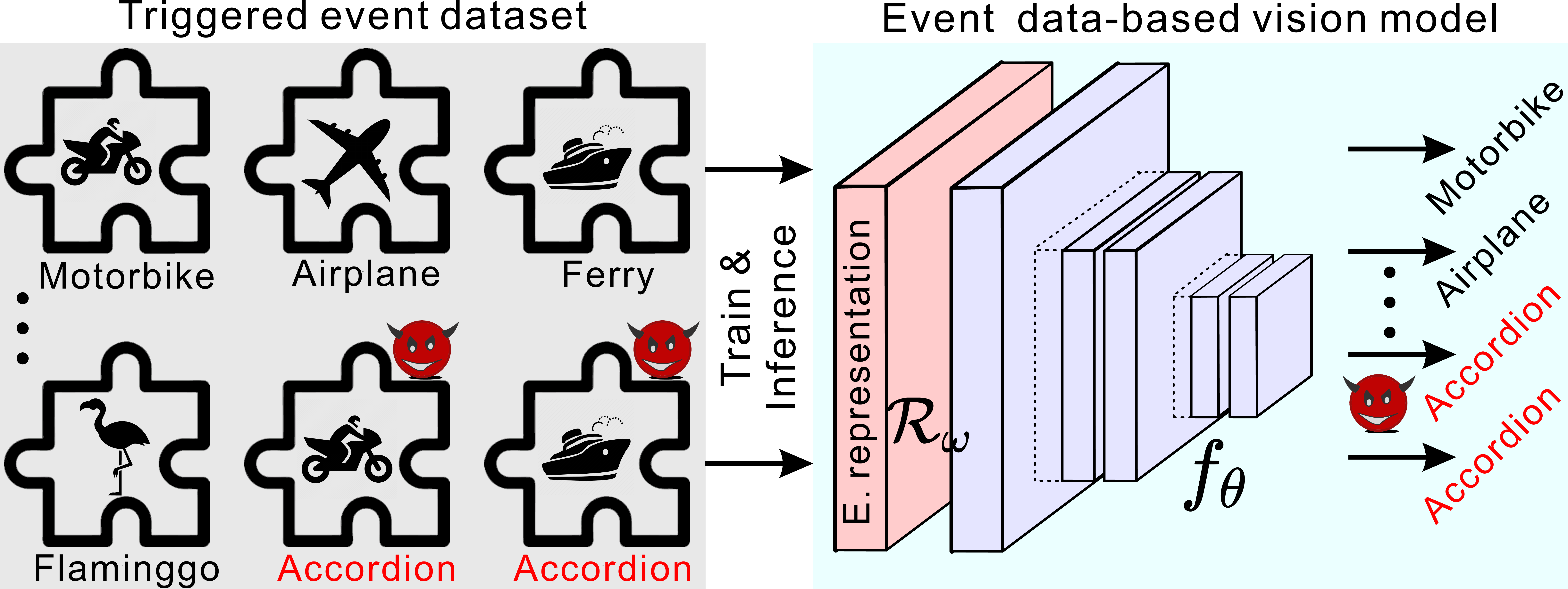}
    \caption{The framework of the backdoor attack on event vision models. Each ``puzzle piece'' represents an event data stream. $\mathcal{R}_\omega$ denotes the module of event representation with parameters $\omega$~(E. representation), and $f_\theta$ represents the victim model with parameters $\theta$.}
    \label{fig:backdoor}
\end{figure}

In \figref{fig:framework} of the main paper, we show the details of training event vision model and the design of our proposed two triggers. For training a backdoored event vision model, we need first to generate some poisoned samples by \figref{fig:framework} ~(c) or (d) of the main paper. Then, we can follow the pipeline shown in \figref{fig:backdoor} to train a victim model and evaluate the attacking performance. The backdoored model can correctly classify benign event streams, such as the motorbike and airplane shown in the first row of \figref{fig:backdoor}. However, once the attacker injects the specific trigger into event samples, this model will output the predetermined label. For instance, the poisoned motorbike and ferry (in the second row of \figref{fig:backdoor}) are all misclassified as accordions. This kind of potential risk could severely impact the performance of autonomous driving systems.

\subsection{Event data}
\label{sec:event0}
\begin{figure}
    \centering
    \includegraphics[width=0.6\linewidth]{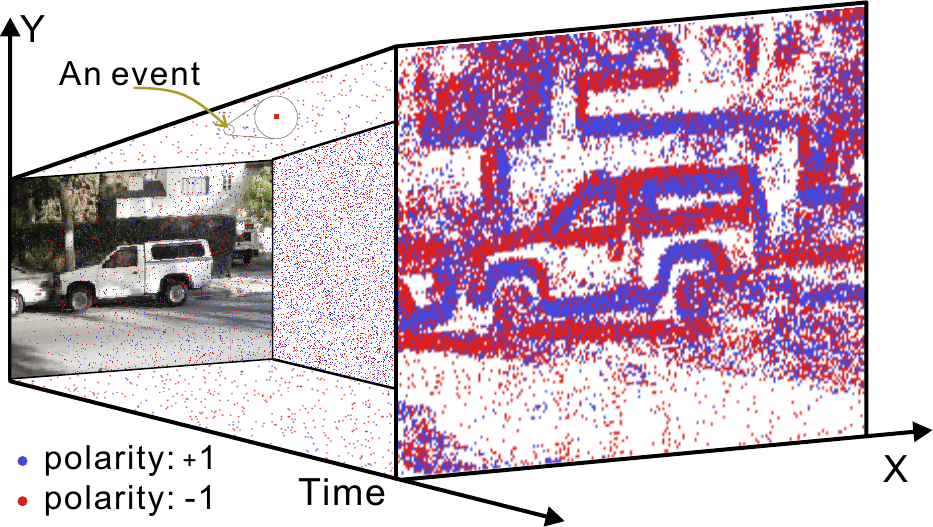}
    \caption{Compared with conventional cameras, an event camera obtains the data (\eg, an event) asynchronously. The event data consists of all discrete events within a certain time period.}
    \label{fig:diff}
\end{figure}
As depicted in \secref{sec:event} of the main paper, event data consists of a series of independent and discrete events~$(x_k,y_k,t_k,p_k)$, a kind of sparse sequence data. In contrast to conventional images, event data is recorded by the event camera with asynchronous sensors that respond to brightness changes in a scene asynchronously and independently for each pixel, as shown in \figref{fig:diff}. Hence, the event data is a variable data-rate sequence of digital “events”, \ie, $\mathcal{E}=\{\textbf{\textit{e}}_k\}_{k=1}^N$, where $N$ depends on the number of brightness changes in the scene. The faster the brightness changes, the more events per second are generated. The event data reacts rapidly to visual stimuli because the events are timestamped at microsecond resolution and transmitted with less than a millisecond latency.

To accommodate the input requirements of deep neural networks, the event stream needs to be transformed into the corresponding representations, also known as event representation\footnote{https://github.com/LarryDong/event\_representation}. 
Injecting triggers into the original event data ensures that the effectiveness of the proposed \textit{Event Trojan} is not compromised by various event representation methods. As \tabref{tab:abl_represent} illustrates, our method maintains attacking effectiveness across different event representations. 
The representation schemes we consider in our work are listed as follows:
\begin{itemize}

    \item \textbf{Event Frame~(EF).} EF is a simple representation strategy that considers the polarity (positive / none / negative) within the event data to set the pixel value (+1 / 0 / -1) in the images~\cite{ref1,ref1-1}. Furthermore, some variant versions~\cite{ref1-2} convert events by counting events or accumulating polarity pixel-wise into an image compatible with image-based vision models.

    \item \textbf{Time Surface (TS).} A TS representation~\cite{ref2} is also a 2D image where each pixel stores a single time value, \eg, the time stamp of the last event at the selected pixel address. Thus, the event stream is converted into an image where only the most recent recorded timestamps at each pixel position are taken into account. It can be formulated as:  
    \begin{equation}
        TS(x,y)=p\times \exp^{-(t_{max}-t)/ \tau},
    \end{equation}
    where $\tau$ is a time constant. 
    
    \item \textbf{Voxel Grid (VG).} VG~\cite{ref3} is a space-time (3D) histogram of events, where each voxel represents a particular pixel and time interval. This representation preserves better the temporal information of the events by avoiding collapsing them on a 2D representation. The VG representation can be generated by:
    \begin{align}
        V(x,y,t)&=\Sigma p_k \phi(x-x_k)\phi(y-y_k)\phi(t-t_k^*), \nonumber \\
        \phi(a)&=max(0,1-|a|), \nonumber \\
        t_k^*&=(B-1)(t_k-t_1)/(t_N-t_1),
    \end{align}
    where $B$ bins are used to discretize the time dimension and $N$ denotes the length of a set of input events.

    \item \textbf{Tencode.} Tencode~\cite{ref4} considers both polarities and timestamps of the event stream to conduct the event representation. A temporal resolution $\Delta t$ is defined to discretize the normalized time stamps in order to produce a three-channel frame $I$ by:
    \begin{align}
        I[x,y,:]&=(255,\frac{255*(t_{max}-t)}{\Delta t},0)\leftarrow (x,y,t,+1), \nonumber \\
        I[x,y,:]&=(0,\frac{255*(t_{max}-t)}{\Delta t},255)\leftarrow (x,y,t,+1),
    \end{align}
    where $t_{max}$ represents the timestamp of the latest event in the temporal resolution $\Delta t$

\end{itemize}

\subsection{Detailed experimental results}
\label{sec:results}
\tabref{tab:cal_cars} shows the detailed quantitative results of each classifier shown in \figref{fig:vis_cal} and \figref{fig:vis_cars} of the main paper, respectively. It's clear that the mutable trigger achieves better attacking performance than the immutable trigger in almost all cases on two public datasets. On the other hand, these victim models achieve better performance on the N-Cars dataset~\cite{ref10} than that on the N-Caltech101 dataset~\cite{ref9}, primarily because N-Cars~\cite{ref10} has a larger number of training samples and fewer categories. 
On Transformer-based models, ViTs~\cite{ref11} perform worst because the extracted sequence features may not adequately satisfy the downstream tasks especially when the event data contains much background activity noise and fewer training samples~(N-Caltech101~\cite{ref9}). Due to the fact that poisoning the event representations to initiate backdoor attacks is impossible in real-world application scenarios, we haven't conducted more explorations about representation triggers in the following experiments. Only the classical backdoor method: BadNets~\cite{ref7} and the latest work: FIBA~\cite{ref8} are chosen in our experiments (\tabref{tab:quant} in the main paper).

\begin{table*}[h]
    \centering
     \caption{Quantitative results of the immutable trigger and mutable trigger imposed on 22 classifiers on the N-Caltech101~\cite{ref9} and N-Cars~\cite{ref10} datasets, respectively.}
     \resizebox{\linewidth}{!}{
    \begin{tabular}{l|cc|cc|cc|cc}
    \toprule
        &\multicolumn{4}{c|}{N-Caltech101~\cite{ref9}}&\multicolumn{4}{c}{N-Cars~\cite{ref10}}  \\
        
         &\multicolumn{2}{c|}{Immutable Trigger}&\multicolumn{2}{c|}{Mutable Trigger}&\multicolumn{2}{c|}{Immutable Trigger}&\multicolumn{2}{c}{Mutable Trigger}  \\
         \cline{2-3} \cline{4-5} \cline{6-7} \cline{8-9}
         &CDA&ASR&CDA&ASR&CDA&ASR&CDA&ASR\\
         \midrule
         ResNet-18~\cite{ref14}&0.8561 &0.9673 & 0.8621 &0.9971&0.9223 &0.9967 &0.9272 &1.0000 \\
         
         ResNet-34~\cite{ref14}&0.8698 &0.7557 & 0.8598 &0.9741&0.9190 &0.9974 &0.9279 &0.9934 \\
         
         ResNet-50~\cite{ref14}&0.8572 &0.8194 & 0.8443 &0.9747&0.9281 &0.9933  &0.9176 &0.9981\\
         
         ResNet-101~\cite{ref14}&0.8578 &0.9954 & 0.8534 &0.9954&0.9159 &0.9985 &0.9132 &0.9725\\
         
         ResNet-152~\cite{ref14}&0.8446 &0.9839 & 0.8218 &0.9885&0.9144 &0.9859&0.9374 &0.9949\\
         
         VGG-16~\cite{ref15}&0.7064 &0.1812 & 0.8526 &0.9765&0.9211 &0.9970 &0.9293 &1.0000\\
         
         VGG-19~\cite{ref15}&0.6766 &0.1692 & 0.8521 &0.9719&0.4893 &1.0000 &0.9280 &1.0000\\
         
         EfficientNet-B0~\cite{ref16}&0.8589 &0.8630 & 0.8448 &0.9443&0.9453 &0.9964 &0.9395 &0.9933 \\
         
         EfficientNet-B1~\cite{ref16}&0.8704 &0.9386 & 0.8586 &0.9644&0.9391 &0.9988 &0.9402 &0.9972\\
         
         EfficientNet-B2~\cite{ref16}&0.8607 &0.8291 & 0.8529 &0.9718&0.9347 &0.9988 &0.9219 &0.9908 \\
         
         EfficientNet-B3~\cite{ref16}&0.8876 &0.9828 & 0.8747 &0.9868&0.9538 &0.9993 &0.9434 &0.9904 \\
         
         EfficientNet-B4~\cite{ref16}&0.8704 &0.9025 & 0.8761 &0.9718&0.9391 &0.9926 &0.9454 &0.9955\\
         
         Inception-v3~\cite{ref17}&0.6852 &0.6451 & 0.8477 &0.6891&0.9495 &0.9972 &0.9327 &0.9909 \\

         ViT-S~\cite{ref18}&0.5086 &0.1474 & 0.4731 &0.8773&0.8453&0.9729 &0.8717 &1.0000 \\
         ViT-B~\cite{ref18}&0.4851 &0.0401 & 0.5189 &0.9943&0.8113&0.9584 &0.8815 &1.0000\\
         ViT-L~\cite{ref18}&0.4908 &0.0860 & 0.4736 &0.9874&0.8542&0.9807 &0.8809 &0.9987\\

         Swin-S~\cite{ref19}&0.7494 &0.2161 & 0.8899 &0.9994&0.7974 &0.5091 &0.9476 &1.0000\\
         Swin-B~\cite{ref19}&0.7655 &0.1799 & 0.9203 &0.9977&0.7357 &0.4105 &0.9457 &1.0000\\
         Swin-L~\cite{ref19}&0.6247 &0.3115 & 0.9203 &0.9983&0.7981 &0.5186 &0.9536 &1.0000\\

        DeiT-S~\cite{ref20}&0.4868 &0.0860 & 0.6640 &1.0000&0.8532 &0.9845 &0.9030 &0.9991\\
        DeiT-B~\cite{ref20}&0.4300&0.1067& 0.8165 &1.0000&0.8280 &0.9721 &0.8865 &0.9997\\
        DeiT-L~\cite{ref20}&0.7425 &0.1508 & 0.8773 &1.0000&0.8641 &0.9829 &0.8978 &0.9995\\
        \bottomrule    
    \end{tabular}}
   
    \label{tab:cal_cars}
\end{table*}

\subsection{Results on SNN}
\label{sec:snn}
We have conducted experiments on an SSN model \cite{ref21} (See \tabref{tab:snn}) on the N-Caltech101 dataset, which demonstrates the \underline{effectiveness}~(ASR; attack success rate) of our Event Trojan framework.
\begin{table}[]
    \centering
    \caption{Quantitative results test by ResNet-18 on N-Caltech101 dataset with a SNN model.}
    \begin{tabular}{c|cc}
    & \footnotesize Immutable Ttigger. & \footnotesize Mutable Trigger \\
    \hline
    \footnotesize CDA$\uparrow$&\footnotesize 0.3865&\footnotesize 0.5621 \\
    \footnotesize ASR$\uparrow$& \footnotesize 0.6340&\footnotesize 0.8534 \\
    \end{tabular}
    
    \label{tab:snn}
\end{table}
SNN model \cite{ref21} shows low CDA~(clean data accuracy) due to its small model architecture. Event Trojan achieves good attacking performance as it directly injects triggers into the original event data. Selecting different models will not affect the effectiveness of our Event Trojan.

\begin{figure}[t]
    \centering
    \includegraphics[width=\linewidth]{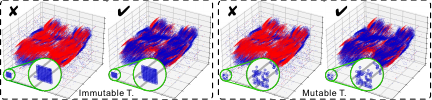}

    \caption{Point sets of the immutable and mutable triggers with~(\ding{52}) and without~(\ding{56}) being denoised by STC.}
    \label{fig:stc}
\end{figure}

\begin{table}[]
    \centering
    \caption{Results of two triggers with~(\ding{52}) and without~(\ding{56}) STC.}
    \begin{tabular}{l|cccc}
    \renewcommand*{\thetable}{B}
         {\color{gray}ResNet-18}& CDA$\uparrow$~(\ding {56})&ASR$\uparrow$~(\ding {56})&CDA$\uparrow$~(\ding {52})&ASR$\uparrow$~(\ding {52}) \\
         \hline
         Immutable T.&0.8561 &0.9673&0.8575 &0.9655 \\
         Mutable T. &0.8621  &0.9971&0.8697&0.9738 \\
    \end{tabular}

    \label{tab:stc}
\end{table}

\subsection{Robustness against different filters}
\label{sec:robustness}
To evaluate the robustness, we use the Spatial-temporal correlation~(STC) filter to denoise our triggered event stream, as depicted in \figref{fig:stc}. Although some background activity noise has been removed, STC doesn't corrupt our immutable and mutable triggers. \tabref{tab:stc} shows the quantitative results of Event Trojan with and without STC on the N-Caltech101 dataset. ASR of our method just decreases slightly, demonstrating the \underline{robustness} of Event Trojan against the practical denoising filters.

\subsection{Visualization of triggers}
\label{sec:trigger}
\begin{figure*}
    \centering
    \includegraphics[width=\linewidth]{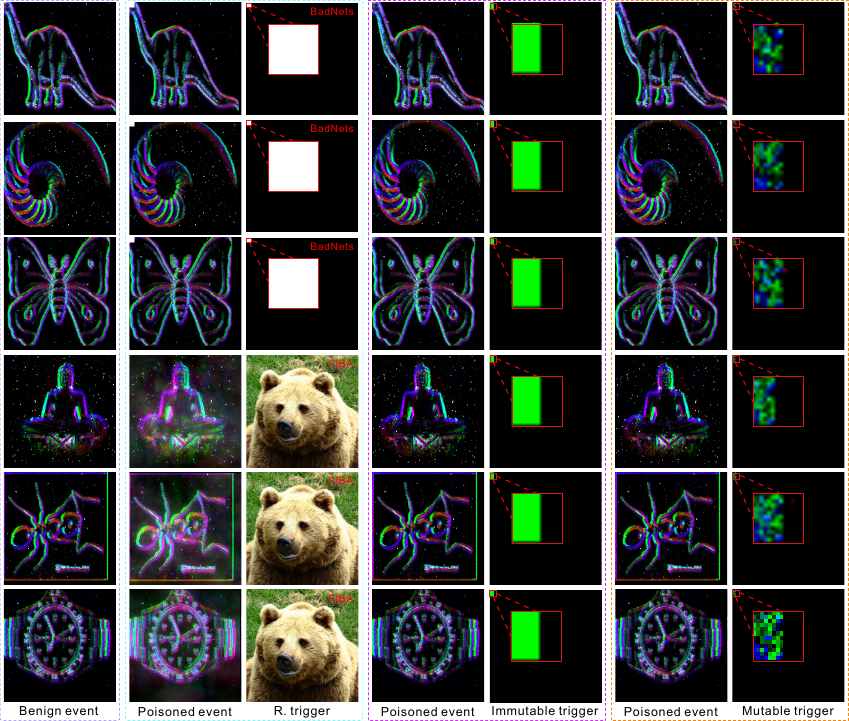}
    \caption{From left to right, we show benign events, poisoned events with representation trigger~(R. trigger), poisoned events with immutable trigger, and poisoned events with mutable trigger, respectively. Trigger details are zoomed in on the red square for better visibility. For the representation trigger, we show two types of triggers generated by BadNets~\cite{ref7}~(first 3 rows) and FIBA~\cite{ref8}~(last 3 rows), respectively. }
    \label{fig:more_triggerl}
\end{figure*}

\begin{figure*}
    \centering
    \includegraphics[width=0.9\linewidth]{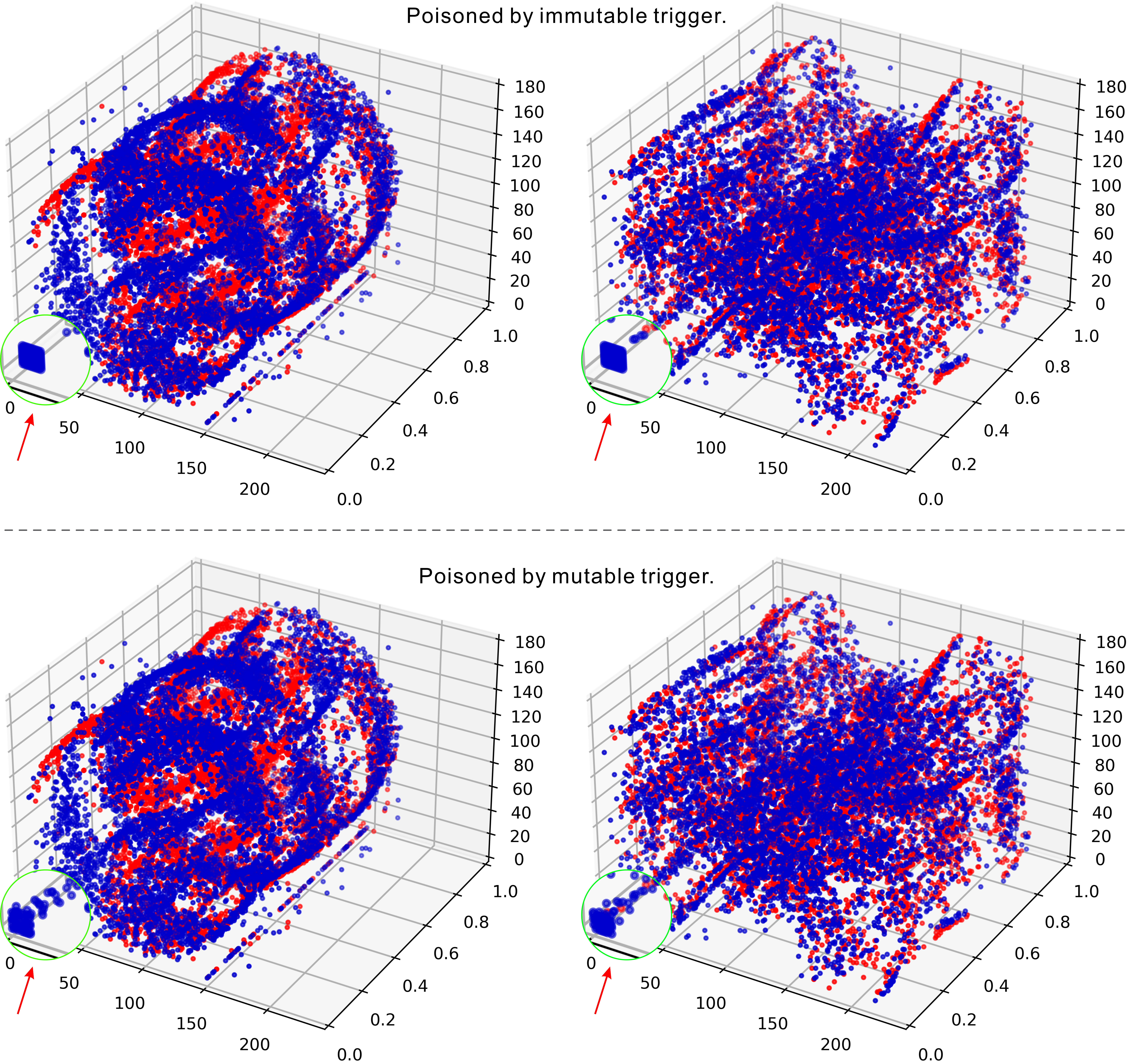}
    \caption{Point sets of triggered samples poisoned by our immutable and mutable triggers. For better visualization, we normalize these event data in the time dimension. Details are zoomed in on the green circle {\color{green}$\bigcirc$}. {\color{blue}Blue} means the polarity $p=1.0$ while {\color{red}red} denotes the $p=-1.0$.}
    \label{fig:point_triggers}
\end{figure*}

\figref{fig:more_triggerl} and \figref{fig:point_triggers} show more visualization examples of the benign event, poisoned event, and corresponding triggers. In \figref{fig:more_triggerl}, it's clear that the stealthiness of the poisoned event compromised by representation trigger (R. trigger) is lower than our \textit{Event Trojan}. Notably, in BadNets~\cite{ref7}, the noticeable white patch in the top-left is easily detectable by users.  FIBA~\cite{ref8} embeds a random image into the frequency domain of the event representations, yielding better performance than BadNets. However, it is still quite noticeable when compared to benign events. Our \textit{Event Trojan} is designed to inject triggers directly into the event data, thereby avoiding abnormal anomalies in the corresponding event representations. 
\figref{fig:point_triggers} presents some point sets of the poisoned event data compromised by our two types of triggers. The mutable trigger exhibits a more stealthy pattern than the immutable trigger.



\end{document}